\documentclass[twocolumn,showpacs,preprintnumbers,superscriptaddress,prl,amsmath,amssymb]{revtex4-1}

\usepackage{graphicx}% Include figure files
\usepackage{dcolumn}% Align table columns on decimal point
\usepackage{bm}% bold math
\usepackage{graphicx}
\usepackage{dcolumn}% Align table columns on decimal point
\usepackage{bm}% bold math
\usepackage{color}
\usepackage{ulem}
\usepackage{gensymb}
\usepackage{braket}
\usepackage{epstopdf}
\usepackage{amsmath}
\usepackage[percent]{overpic}

\begin{document}

\newcommand{\customsection}[1]{\textit{#1.\textemdash}}
\newcommand{\temperaturerange}[3]{#1\,K $\leq$ #2 $\leq$ #3\,K}

\title{High antiferromagnetic transition temperature of a honeycomb compound SrRu$_2$O$_6$}

\author{W. Tian}
\affiliation{Quantum Condensed Matter Division, Oak Ridge National Laboratory, Oak Ridge, Tennessee 37831, USA}

\author{C. Svoboda}

\affiliation{Department of Physics, The Ohio State University, Columbus, OH 43210, USA}

\author{M. Ochi}
\affiliation{RIKEN Center for Emergent Matter Science, 2-1 Hirosawa, Wako, Saitama 351-0198, Japan}

\author{M. Matsuda}
\affiliation{Quantum Condensed Matter Division, Oak Ridge National Laboratory, Oak Ridge, Tennessee 37831, USA}

\author{H. B. Cao}
\affiliation{Quantum Condensed Matter Division, Oak Ridge National Laboratory, Oak Ridge, Tennessee 37831, USA}

\author{J.-G. Cheng}
\affiliation{Beijing National Laboratory for Condensed Matter Physics, and Institute of Physics, Chinese Academy of Sciences, Beijing 100190, P. R. China}

\author{B. C. Sales}
\affiliation{Materials Science and Technology Division, Oak Ridge National Laboratory, Oak Ridge, Tennessee 37831, USA}

\author{D. G. Mandrus}
\affiliation{Materials Science and Technology Division, Oak Ridge National Laboratory, Oak Ridge, Tennessee 37831, USA}
\affiliation{Department of Materials Science and Engineering, University of Tennessee, Knoxville, Tennessee 37996, USA}

\author{R. Arita}
\affiliation{RIKEN Center for Emergent Matter Science, 2-1 Hirosawa, Wako, Saitama 351-0198, Japan}

\author{N. Trivedi}
\affiliation{Department of Physics, The Ohio State University, Columbus, OH 43210, USA}

\author{J.-Q. Yan}
\affiliation{Materials Science and Technology Division, Oak Ridge National Laboratory, Oak Ridge, Tennessee 37831, USA}
\affiliation{Department of Materials Science and Engineering, University of Tennessee, Knoxville, Tennessee 37996, USA}

\date{\today}

\begin{abstract}
We study the high temperature magnetic order in SrRu$_2$O$_6$ by measuring magnetization and neutron powder diffraction with both polarized and unpolarized neutrons. SrRu$_2$O$_6$ crystallizes into the hexagonal lead antimonate (PbSb$_2$O$_6$, space group \textit{P}$\overline{3}$1\textit{m}) structure with layers of edge-sharing RuO$_6$ octahedra separated by  Sr$^{2+}$ ions. SrRu$_2$O$_6$ is found to order at $T_N$=565\,K with Ru moments coupled antiferromagnetically both in-plane and out-of-plane. The magnetic moment is 1.30(2) $\mu_\mathrm{B}$/Ru at room temperature and is along the crystallographic \textit{c}-axis in the  G-type magnetic structure. We perform density functional calculations with constrained RPA to obtain the electronic structure and effective intra- and inter-orbital interaction parameters. The projected density of states show strong hybridization between Ru 4$d$ and O 2$p$. By downfolding to the target $t_{2g}$ bands we extract the effective magnetic Hamiltonian and perform Monte Carlo simulations to determine the transition temperature as a function of inter- and intra-plane couplings. We find a weak inter-plane coupling, 3\% of the strong intra-plane coupling, permits three-dimensional magnetic order at the observed $T_N$.

\end{abstract}


\maketitle

%\section{Introduction}

It is evident that for the design of the next generation of multifunctional devices, we need new paradigms, new principles and new classes of materials. Magnetism is arguably the most technologically important property arising from electron interactions.
One of the central questions has been how to create magnetic materials with high transition temperatures $T_c$ for room temperature devices. Broadly two paradigms define the formation of the magnetic state starting with fermions at finite temperatures: for weak Coulomb interactions $U/W$ compared to the bandwidth, one expects a Fermi liquid at finite temperatures, followed by a Fermi surface nesting instability that opens a gap in the spectrum resulting in a Slater antiferromagnet below $T_c\sim W e^{-c\sqrt{W/U}}$ which is exponentially suppressed in the coupling ($c$ is a constant). In the opposite regime for $U/W\gg 1$, local moments form on a much higher temperature scale $T^\ast \approx U$ opening a large Mott gap and order on the scale of antiferromagnetic (AF) superexchange $J\sim W^2/U$.
The transition temperature as a function of $U/W$ reaches its maximum in the fluctuating regime between these two regimes. SrTcO$_3$ is currently believed to be at the maximum with a $T_c\approx 1000$ K~\cite{RodriguezSrTcO3,MravljeSrTcO3,Borisov,Franchini,Middey,Gordon}.  Tuning $U/W$ by combining different 3$d$ and 5$d$ ions in double perovskites also increases $T_c$ well above room temperature~\cite{Nandini}, as in Sr$_2$CrReO$_6$ with a $T_c=635$ K~\cite{KatoSr2CrReO6} and Sr$_2$CrOsO$_6$ with a $T_c=720$ K~\cite{KrockenbergerSr2CrOsO6}. These are examples of Mott-Hubbard antiferromagnetic insulators. NaOsO$_3$ orders at $T_c = 410$\,K~\cite{ShiNaOsO3,CalderNaOsO3}, which is a rare example of Slater insulator approaching the fluctuating region from the itinerant side. All the above reported high $T_c$ compounds have perovskite structures with a $d^3$ electronic configuration.
%The above qualitative description is substantiated below by a more detailed understanding of the role played by inter- and intra- orbital Coulomb interactions and hybridizations within the unit cell between transition metal \textit{d} and oxygen \textit{p} orbitals as well as between unit cells.

Recently, a metastable compound SrRu$_2$O$_6$ with a quasi two dimensional structure was proposed to order antiferromagnetically with $T_N$ above 500 K\cite{HileySrRu2O6}. SrRu$_2$O$_6$ crystallizes into the hexagonal lead antimonate (PbSb$_2$O$_6$) structure with the space group \textit{P}$\overline{3}$1\textit{m}. As shown in the inset of Fig.\,1, the structure consists of layers of edge-sharing RuO$_6$ octahedra separated by  Sr$^{2+}$ ions sitting in the oxygen octahedral interstices. In the \textit{ab} plane, the Ru ions form a honeycomb array. In the previous study\cite{HileySrRu2O6}, the magnetization was measured up to 500 K without finding a signature for the magnetic transition, though
the room temperature neutron powder diffraction observed extra reflections absent in x-ray measurements. In this Letter, we report our magnetic and neutron diffraction study of SrRu$_2$O$_6$ up to 750 K. Our diffraction measurements with both polarized and unpolarized neutrons confirm that SrRu$_2$O$_6$ orders antiferromagnetically at $T_N=565$ K with a magnetic moment of 1.30(2)\,$\mu_\mathrm{B}$/Ru along $c$-axis. The magnetic measurements suggest that strong two-dimensional magnetic correlations persist above $T_N$ as highlighted in Fig.\,1 above the kink. Our band structure calculations show strong Ru 4\textit{d} and O 2\textit{p} hybridization, which accounts for the reduced moment of Ru$^{5+}$ ions and possibly the high magnetic ordering temperature as in SrTcO$_3$. We analyze the different exchange pathways and derive an effective magnetic model with predominantly antiferromagnetic interactions within the plane,
with exchange parameters determined from downfolding and constrained RPA, as well as antiferromagnetic interactions between planes. Using Monte Carlo simulations of such a coupled layer model we calculate the dependence of $T_N$ on interlayer coupling and by comparison with the experimental $T_N$
determine the interlayer magnetic interaction to be small, about 3\% of the in-plane interaction.

\begin{figure} \centering \includegraphics [width = 0.47\textwidth] {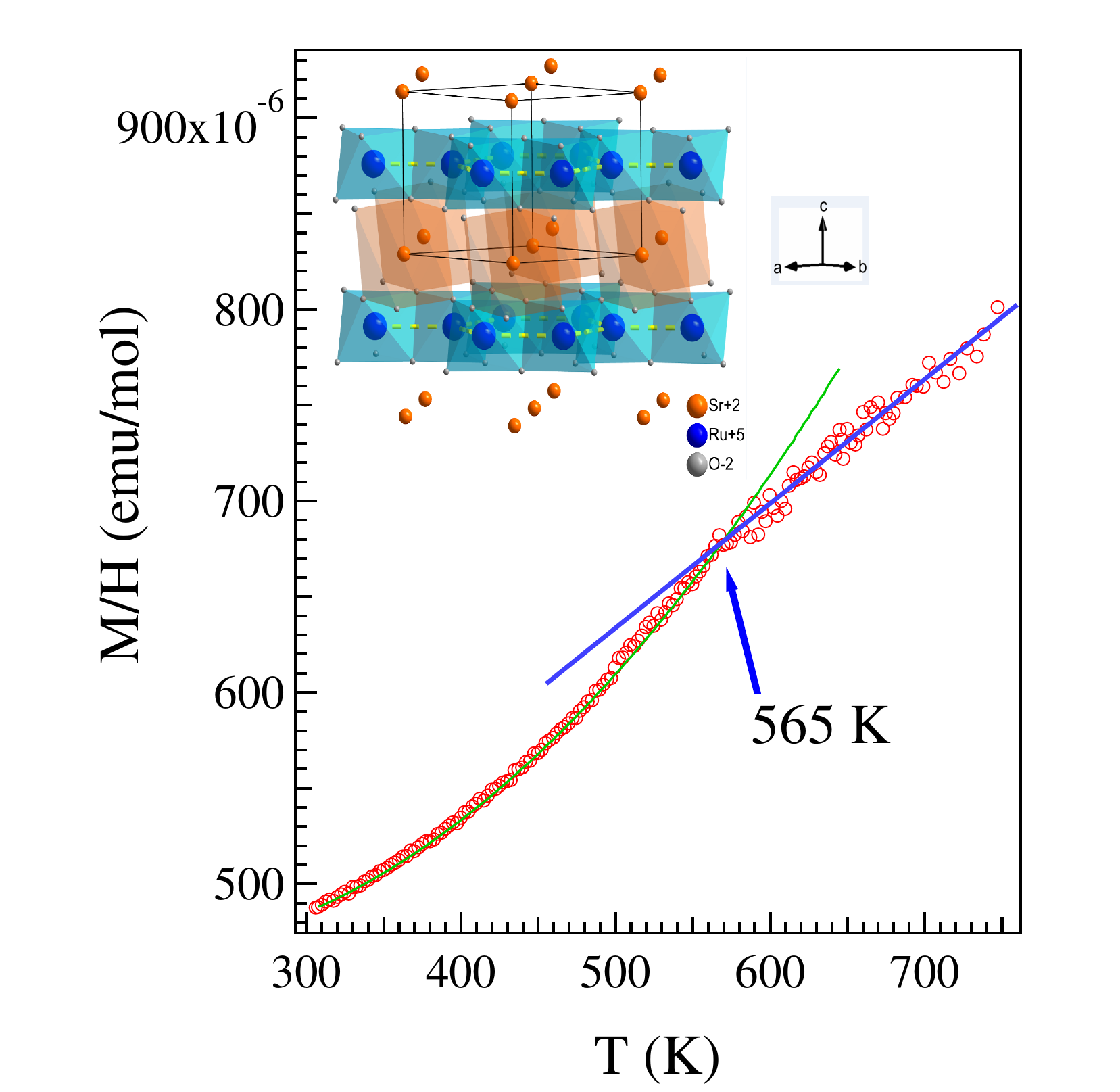}
\caption{(color online) Magnetic susceptibility in the temperature range \temperaturerange{300}{T}{750} measured upon cooling in an applied field of 50\,kOe. The solid curves highlight the slope change at 565\,K. Inset shows the crystal structure. }
\label{Mag-1}
\end{figure}

\begin{figure} [ht!]
\centering \includegraphics [width = 0.47\textwidth] {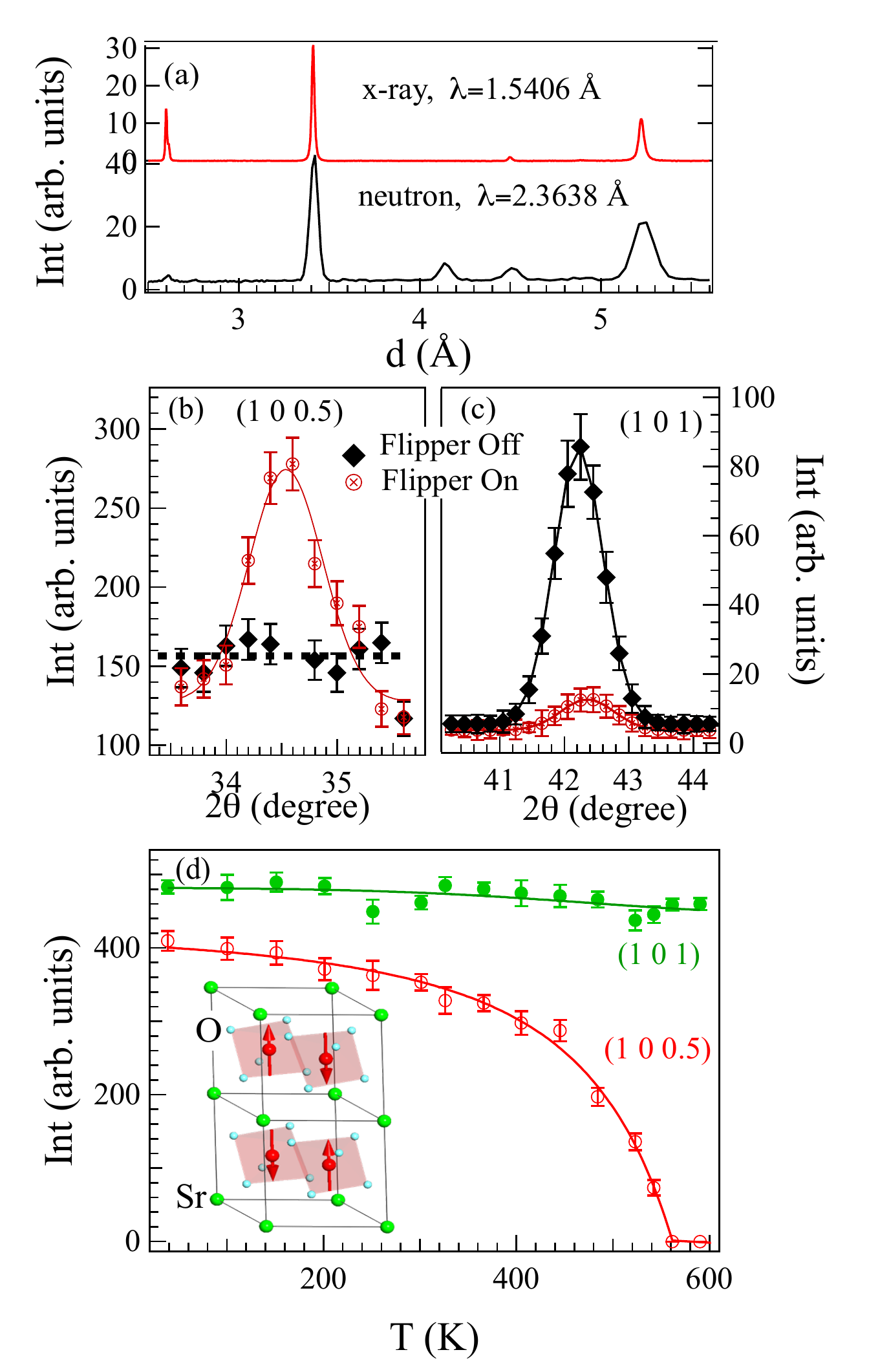}
\caption{(color online) (a) Room temperature x-ray and neutron powder diffraction patterns in a narrow d-range highlighting the extra reflections observed by neutron diffraction. (b) and (c) show the (1 0 0.5) magnetic and (1 0 1) nuclear peaks measured with the polarized neutron in the horizontal field configuration P$_0$ $\parallel$ Q at room temperature, respectively. The observed weak intensity in (c) with flipper on comes from the finite instrumental flipping ratio which we estimate to be ~1/10 by comparing the integrated intensity of the ($-+$) and ($++$) scans of (1 0 1). (d) The temperature dependence of integrated intensity of (1 0 0.5) and (1 0 1) peaks. The solid curves are a guide to the eye. Inset shows the G-type magnetic structure.}
\label{Fig2-1}
\end{figure}

We synthesize polycrystalline SrRu$_2$O$_6$ by a hydrothermal technique as reported previously\cite{HileySrRu2O6} and confirm that the obtained powder is a single phase from room temperature x-ray powder diffraction.  All the peaks observed can be indexed with the space group \textit{P}$\overline{3}$1\textit{m} and lattice parameters $a$\,=\,5.2037(2)\,$\mathrm{\AA}$ and $c\,=\,5.2341(4) \mathrm{\AA}$.  Fig.\,1 shows the temperature dependence of magnetic susceptibility measured in the temperature range \temperaturerange{300}{T}{750} using a Quantum Design magnetic property measurement system. The  data collected in warming and cooling processes overlap suggesting little or no sample decomposition below 750\,K. Above room temperature, the magnetic susceptibility increases with increasing temperature. As highlighted by the solid curves, there is a slope change at 565\,K, which signals possible long range magnetic order.

To study the nature of this slope change, we carry out neutron diffraction experiments on about 0.4 g powder sample in the temperature range \temperaturerange{40}{T}{600} using the HB-1A and HB1 triple axis spectrometers and HB-3A four-circle spectrometer located at the High Flux Isotope Reactor at Oak Ridge National Laboratory~\cite{supportingMaterials}. As shown in Fig.\,2(a), neutron powder diffraction observed reflections that are absent in x-ray powder diffraction pattern. To confirm the magnetic origin of these extra reflections, we performed diffraction measurement using polarized neutrons with the neutron polarization parallel to the moment transfer: $\mathrm{P}_0\parallel \mathrm{Q}$. With the spin flipper off or on, we measured both the ($++$) non-spin-flip and the ($-+$) spin-flip scattering of different reflections of interest. Fig.\,2(b) and (c) show the intensity of (1 0 0.5) and (1 0 1), respectively. As discussed in Refs. \cite{polarized1,polarized2,polarized3}, coherent nuclear scattering is always ($++$) non-spin-flip scattering because it never causes a reversal or spin flip of the neutron spin direction upon scattering. On the other hand, magnetic scattering depends on the relative orientation of the neutron polarization P$_0$ and the scattering vector Q. Only those spin components which are perpendicular to the scattering vector are effective. Thus for a fully polarized neutron beam with the horizontal field configuration, $\mathrm{P}_0\parallel \mathrm{Q}$, all magnetic scattering is ($-+$) spin-flip scattering, and ideally no ($++$) non-spin-flip scattering will be observed. The strong intensity shown in Fig. 2(c) was observed in the ($++$) channel as expected for the nuclear (1 0 1) peak. The strong scattering detected in the ($-+$) spin-flip channel confirmed the magnetic origin of the (1 0 0.5) peak.

The diffraction study with polarized neutrons clearly shows that the extra reflections come from a long range magnetic order instead of any structural transition.  Those extra reflections can be indexed on the basis of the magnetic scattering with a G-type AF structure with the magnetic moment direction along the crystallographic $c$-axis. The Rietveld refinement of the neutron diffraction patterns~\cite{supportingMaterials} yields a magnetic moment of 1.30(2) $\mu_\mathrm{B}$/Ru at room temperature. The magnetic moment increases slightly to 1.34(3) $\mu_\mathrm{B}$/Ru upon cooling to 40 K. The moment is smaller than the spin moment of 3\,$\mu_{B}$ expected for a a half-filled $t_{2g}$ band, which signals a strong covalency of the Ru-O bonds.

We further follow the temperature dependence of the integrated intensity of the (1 0 0.5) magnetic peak in the temperature range of \temperaturerange{40}{T}{600}. As shown in Fig.\,2(d), the (1 0 0.5) magnetic peak disappears around 565 K, where a slope change is observed in Fig.~1 in the temperature dependence of the magnetic susceptibility. Figure\,2(d) also shows the evolution with temperature of the integrated intensity of the (1 0 1) nuclear peak, which shows little temperature dependence in the temperature ranged studied. The coincidence of the disappearance of (1 0 0.5) magnetic peak and the slope change in magnetic susceptibility suggests that a G-type long range magnetic order takes place at 565 K in SrRu$_2$O$_6$.

In SrRu$_2$O$_6$ the Ru ions are in a $d^3$ electronic configuration in edge-shared octahedral cages formed by the O atoms. It is important to note that the Ru-O-Ru bond angle is close to 90\degree\ hence both ferromagnetic (F) and AF mechanisms through intermediate oxygens are active in fourth order processes according to Goodenough-Kanamori rules~\cite{Goodenough, Kanamori}. There are three competing processes that contribute to the exchange interaction:  (a) the direct overlap of the half filled $t_{2g}$ orbitals produces a second order AF interaction. (b) The transfer of electrons between an oxygen $p_z$ orbital and Ru $d_{zx}$ and $d_{yz}$ orbitals on two neighboring Ru atoms results in an AF superexchange coupling. (c) The transfer of electrons between Ru $t_{2g}$ orbitals and mutually orthogonal oxygen $p$ orbitals results in a F interaction driven by Hund's coupling on oxygen.

It is rather intriguing that SrRu$_2$O$_6$ orders at such a high temperature, given the competing magnetic interactions. To estimate the relative magnitude of the above competing interactions and to understand the mechanism for the high AF ordering temperature, we perform DFT calculations including constrained RPA to obtain effective hopping and interaction parameters.  We also perform spin density functional calculation with the {\sc wien2k}~\cite{wien2k} package using the exchange-correlation functional proposed by Perdew {\it et al.}~\cite{PW92}
and the full-potential linearized augmented plane-wave method including the spin-orbit coupling. In the calculation, we use the experimental lattice parameters and atomic configurations determined at room temperature~\cite{HileySrRu2O6}.

\begin{figure}[ht]
\centering
\includegraphics[width=8.5cm]{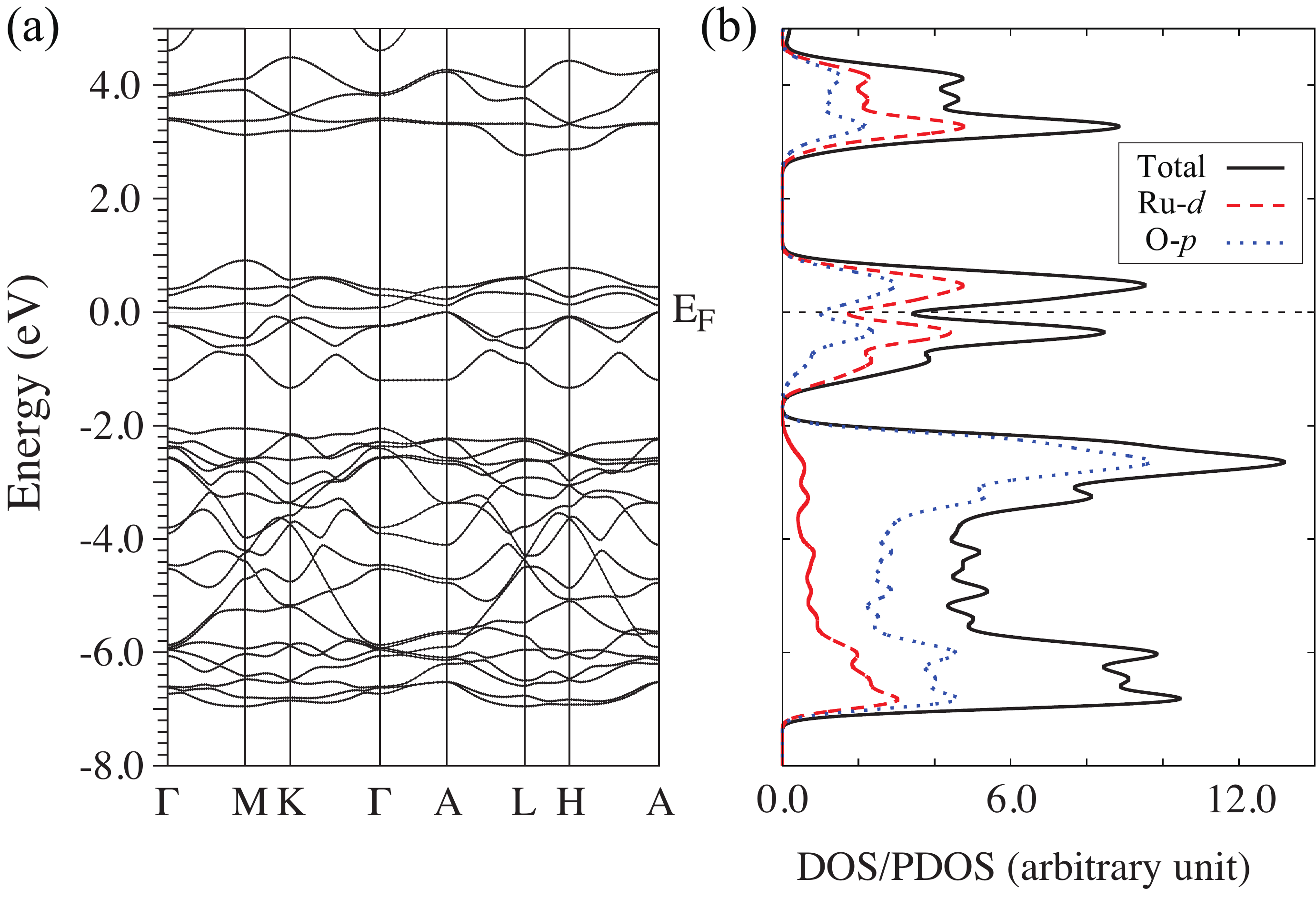}
\caption{(color online):
(a) Band dispersion and (b) (projected) density of states for the non-magnetic state.
}
\label{LDA}
%}
\end{figure}

Our DFT calculation shows that the G-type AF state is the most stable. The non-magnetic (NM) and the C-type AF (i.e., AF in the $ab$-plane and F along the $c$-axis) states have higher energy than that for the G-type AF state, and there is no ferromagnetic or A-type AF (i.e., F in the $ab$-plane and AF along the $c$-axis) metastable solution. These results indicate that there is a strong in-plane AF correlation compared with that for the out-of-plane direction. Calculated local spin magnetic moment of Ru atoms is about 0.9 $\mu_\mathrm{B}$ per atom for the G-type AF state and in reasonable agreement with our experiment. Fig.~\ref{LDA} shows the band dispersion and (projected) density of states (DOS) for the NM state. The $t_{2g}$ bands~\cite{t2g_crystal} are well isolated from the $e_g$ and oxygen $p$ bands. Strong Ru-O hybridization is seen in the DOS, which points to the origin of the reduced local spin magnetic moment of Ru atoms.

Using a density response code~\cite{Anton} recently developed for the Elk branch of the original {\sc exciting fp-lapw} code~\cite{elk}, we derive low-energy effective models. Starting with the DFT calculation for the NM state,
we first construct the Wannier functions and calculate the transfer integrals between them.
Next, we evaluate the interaction parameters, the Coulomb repulsion $U({\bf r})$ and the exchange coupling $J({\bf r})$
by the constrained RPA~\cite{Aryasetiawan}. For simplicity, we neglect the spin-orbit interaction since it is irrelevant in the following analysis.

To evaluate the AF coupling through the direct overlap between $d$ orbitals, we derive a low-energy effective model for the Ru $t_{2g}$ and O $p$ bands in the energy window [$-7.0$:$+1.0$] eV. We obtain the onsite Hubbard $U_{d} = 5.3$ eV and the largest transfer hopping between $d$ orbitals $t_{dd} = 0.19$ eV, which result in a small AF coupling, $J\sim 4t_{dd}^2/U_{d} = 0.03$ eV.

We also derive an effective model only for the Ru $t_{2g}$ bands in the energy window [$-1.4$:$+1.0$] eV.
In this model, the Wannier functions are the Ru $t_{2g}$ orbitals hybridized with the surrounding O $p$ orbitals,
which allows a direct evaluation of the superexchange couplings.
We obtain the on-site Hubbard $U=U({\bf r}=0) = 2.7$ eV, the Hund's coupling $J_H=J({\bf r}=0) = 0.28$ eV,
the nearest-neighbor off-site Coulomb interaction $V=1.1$ eV,
and the largest nearest-neighbor transfer hopping $t = 0.28$ eV.
For the AF superexchange coupling, these values result in $J_{AF}\sim 4t^2/(U-V)=0.20$ eV.
On the other hand, the F superexchange coupling $J_F$ is evaluated as the nearest-neighbor off-site direct exchange $\sim$ $0.03$ eV. The superexchange AF coupling $J_{AF}$ dominates over $J$ and $J_F$. These estimates are replaced by an exact treatment in the following analysis.
%Note that the $J$ values defined here for a one pair of orbitals do not equal those defined for an $S=3/2$ spin.

In the atomic limit, each site has an $S=3/2$ spin. When the two sites are coupled, the eigenstates can be labeled by the total spin $S=0,1,2,3$. Using the effective values of $t$, $U$, and $J_H$ in the $t_{2g}$ effective model~\cite{supportingMaterials}, we perform exact diagonalization for two sites to obtain the energies of states labeled by $S$.  These eigenvalues determine the exchange constants of a general effective spin-$3/2$ Hamiltonian for two sites as $H_{\mathrm{eff}}= E_0+ J_1 (\vec{S}_1 \cdot \vec{S}_2) + J_2 (\vec{S}_1 \cdot \vec{S}_2)^2 + J_3 (\vec{S}_1 \cdot \vec{S}_2)^3$.
We find $J_1 = 45.6$ meV, $J_2 = -2.0$ meV, and $J_3 = 0.5$ meV.
The interaction at the two site level is then primarily Heisenberg AF.

\begin{figure}
\centering
\begin{overpic}[width=8.6cm]{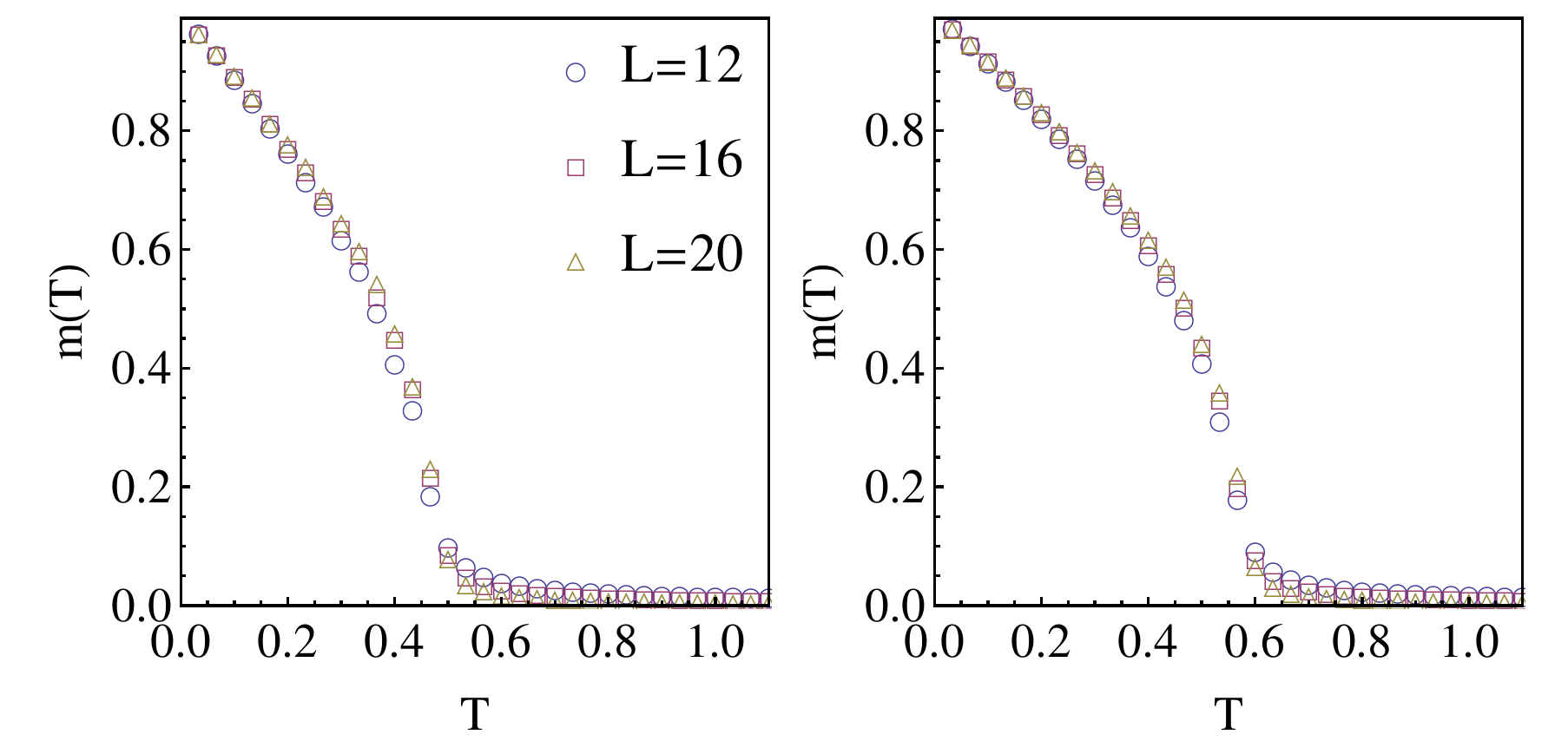}
% \put (15,59) {(a)}
% \put (65,59) {(b)}
 \put (13,12) {(a)}
 \put (61,12) {(b)}
\end{overpic}
\caption{
(color online).
Classical Monte Carlo simulations using the model of \eqref{twoTermHeisenbergModel} showing the
staggered magnetization per site versus temperature in units of $J_{\parallel}$ for (a) $J_{\perp} / J_{\parallel} = 0.03$ and (b) $J_{\perp} / J_{\parallel} = 10^{-1}$ for three system sizes $L$ (total system size $4L^3$). Lowering $J_{\perp}$ at fixed $J_{\parallel}$ reduces the transition temperature.
The transition temperatures extracted by finite size scaling are $0.49 J_{\parallel}$ and $0.59 J_{\parallel}$ respectively.
}
\label{finitesize}
\end{figure}

For purposes of modeling the system, we retain the Heisenberg AF nearest neighbor interactions within the plane ($J_{\parallel}$, 3 neighbors) taken as $J_1$. Also, motivated by the experimental observation of G-type ordering, we introduce, in addition, a coupling between nearest neighbor planes ($J_{\perp}$, 2 neighbors). The classical Hamiltonian describing magnetism in SrRu$_2$O$_6$ is given by:
\begin{equation}
H_{\mathrm{classical}} = J_{\parallel} \sum_{\langle i j \rangle \parallel} \vec{S}_i \cdot \vec{S}_j + J_{\perp} \sum_{\langle i j \rangle \perp} \vec{S}_i \cdot \vec{S}_j
\label{twoTermHeisenbergModel}
\end{equation}
For $J_{\perp}=0$, according to the Mermin-Wagner theorem the long wave length spin waves destroy the long range order and consequently, the transition temperature tends to zero in the thermodynamic limit. However, a small inter-plane coupling can stabilize magnetic order. Our DFT calculations show that the inter-plane hopping parameters are small with respect to the in-plane parameters ($t_{\perp} / t_{\parallel} \sim 0.1$) which suggest that the AF inter-plane couplings $J_{\perp}$ are between $0.1 J_{\parallel}$ and $0.01 J_{\parallel}$. We perform classical Monte Carlo simulations on a layered honeycomb lattice to obtain the transition temperature as a function of the ratio of these coupling constants.
Fig.\,\ref{finitesize} gives the results of these simulations for two values of $J_{\perp}/J_{\parallel}$.

The transition temperature obtained from Monte Carlo $T_N^{\mathrm{MC}} (J_\parallel,J_\perp)$ depends on the in-plane and out-of-plane couplings in general
as seen in Fig.\,\ref{finitesize}.
Using our estimated value for  $J_1$ obtained from downfolding and exact diagonalization, we obtain the AF Heisenberg spin-3/2 coupling constant
$J_{\parallel} = (3/2)^2 J_1= 102.6$ meV or $1190$ K.
Further by using the experimental transition temperature of 565 K, we find that $J_\perp=36$ K or equivalently $J_{\perp}/J_{\parallel} \approx 0.03$
fits the experimental results. The theoretical results also suggest that the transition temperature can be enhanced by
increasing the in-plane coupling, for example by chemical or applied pressure.

As illustrated in the inset of Fig.\,1, the quasi two dimensional crystal structure of SrRu$_2$O$_6$ distinguishes itself from other reported high $T_c$ compounds with a perovskite structure. With the nonmagnetic Sr spacing layers the inter-plane coupling is expected to be weak. This is supported by our Monte Carlo simulations. The inter-plane coupling is small, but critical for the three dimensional magnetic order. The strong in-plane magnetic interaction is dominated by the AF superexchange coupling between rutheniums mediated by oxygen. We notice in Fig.\,1 only one weak slope change around $T_N$ and also the magnetic susceptibility increases linearly with increasing temperature above $T_N$. The absence of a Cure-Weiss-like paramagnetic behavior above $T_N$ suggests that strong two dimensional magnetic fluctuations exist above $T_N$ and persist until this compound decomposes around 800\,K.

Our theoretical modeling highlights the mechanism for the large AF temperature. Similar to the mechanism for the perovskite SrTcO$_3$ in Ref.~\cite{MravljeSrTcO3}, we find that
SrRu$_2$O$_6$ is close to the fluctuating regime with \textit{U}$\sim$\textit{W} on the localized side. This is facilitated by substantial hybridization between Ru 4$d$ and O 2$p$
within the unit cell as seen from the projected density of states in Fig.\,3(b) that reduces the effective $U$. In addition there is large hybridization between unit cells
generating a large band width $W$, bringing this material close to the crossover region where $T_N$ is enhanced. Thus, the large covalency reduces the Ru moment but also facilitates a strong (in-plane) AF superexchange interaction important for the high $T_N$.

With multiple $t_{2g}$ orbitals and Coulomb correlations that generate magnetism on a honeycomb lattice, it is possible that this material has interesting topological properties that still need to be explored, especially as we replace Ru with the heavier Os and spin-orbit coupling becomes important. SrRu$_2$O$_6$, therefore,  provides a new materials platform for studying the mechanism inducing high temperature magnetic order and other exotic phenomena in 4$d$ and 5$d$ transition metal oxides.

Work at ORNL was supported by the U.S. Department of Energy, Office of Science, Basic Energy Sciences, Materials Sciences and Engineering Division and Scientific User Facilities Division. The theoretical modeling (CS and NT) and part of the materials synthesis were supported by the CEM, and NSF MRSEC, under grant DMR-1420451. RA thanks a fruitful discussion with S. Sakai and Y. Nomura.


\begin{references}
\bibitem{t2g_crystal}
We call the Ru-$d$ bands near the Fermi energy the ``$t_{2g}$'' bands for convenience whereas the $t_{2g}$ bands experience an additional crystal field splitting in this material.
However, it is just a unitary transformation of the basis in our model, and so does not affect the following analysis.
\end{references}
\end{document}